\documentclass{article}

   \PassOptionsToPackage{numbers, compress}{natbib}

\usepackage[preprint]{neurips_2022}

\usepackage[utf8]{inputenc} 
\usepackage[T1]{fontenc}    
\usepackage{hyperref}       
\usepackage{url}            
\usepackage{booktabs}       
\usepackage{amsfonts}       
\usepackage{nicefrac}       
\usepackage{microtype}      
\usepackage{xcolor}         
\usepackage{graphicx}       
\usepackage{float}

\usepackage{natbib}
\title{Zero-day DDoS Attack Detection}

\author{%
  Cameron Boeder \\
  MLSP Program\\
  University of Wisconsin Madison\\
  Madison, WI 53706\\
  \texttt{cboeder@wisc.edu} \\
  \And
  Troy Januchowski\\
  MLSP Program\\
  University of Wisconsin Madison\\
  Madison, WI 53706 \\
  \texttt{tjanuchowski@wisc.edu} \\
}

\begin{document}

\maketitle

\section{Abstract}

The ability to detect zero-day (novel) attacks has become essential in the network security industry. Due to ever evolving attack signatures, existing network intrusion detection systems often fail to detect these threats. This project aims to solve the task of detecting zero-day DDoS (distributed denial-of-service) attacks by utilizing network traffic that is captured before entering a private network. Modern feature extraction techniques are used in conjunction with neural networks to determine if a network packet is either benign or malicious.

\section{Introduction}

\subsection{Problem Description}

Network security has become more important over the last decade due to the increasing number of new attack techniques being used against computer networks. Many of these attacks are attempts to compromise a computer network by overwhelming the network with information, which can often lead to devices slowing down, shutting down, or opening the network to new vulnerabilities. These types of attacks are commonly referred to as distributed denial-of-service attacks (DDoS), and the technique for generating this type of attack has been evolving for the past twenty years~\cite{cloudflare1}~\cite{imperva}. Intrusion detection systems (IDS) are used to detect DDoS attacks on a network so that actions can be taken to prevent the network from becoming unavailable ~\cite{hung-jenliao}.

It is difficult to detect newer types of DDoS attacks that a model has not been trained on before. One of the main hurdles to this problem is feature extraction: how can useful information from an abundance of network traffic be extracted. Raw network traffic, such as a pcap or tcpdump file, can often be messy and unintuitive to work with. This project surveys different feature extraction techniques, both manual and automatic, and dataset scenarios. These feature extraction techniques and dataset combinations were then tested using a selection of machine learning classification models. 

\subsection{History and Importance}

DDoS attacks have affected many individuals and organizations over the past several decades. A typical result of the attack involves shutting down a home network or a company's website. In more extreme cases, these attacks have been used to make servers vulnerable to further attacks, or have been used to hold a  company’s website at ransom until a payment is made~\cite{imperva}~\cite{cloudflare2}. Using a machine learning-based IDS can prevent the DDoS attack from having a significant effect on the targeted network. If caught early enough, equipment, such as a network switch, can be shut off to prevent any issues with the hardware connected to the network. In the case of an online business, this could end up saving both time and money. DDoS attacks have also been used, in combination with other attacks, to steal information. An added benefit of catching these attacks early would be to prevent such information leaks from happening. Early DDoS attacks in the late 1990s were mainly volumetric attacks, where a network is overwhelmed with data to the point that the network stops working. Attack types have evolved over the last several decades to also include attacks that focus on different network protocol types, such as HTTP and TCP. These types of attacks can often be slower and harder to detect.~\cite{A10vol}~\cite{A10prot} 

There are existing tools today that can prevent this type of attack from affecting the target network, but these can often require the purchase of expensive equipment. There are also existing machine learning models available to detect attacks, but many of the models are tested on older datasets and attack methods. Additionally, many of these models do not utilize machine learning for feature extraction. In cases of smaller companies that are not using machine learning, this type of detection may be based on manually sorted data by IT personnel, which can be error-prone and arduous. This project  focuses on training a network that would be able to identify new types of DDoS attacks, and on both automated and manual feature extraction.  Ideally, the output of this project will be an effective, cheap solution to preventing a DDoS attack from affecting a network, especially for smaller businesses with limited resources.

\subsection{Related Work}

To further motivate this project, existing solutions not utilizing machine learning techniques were first studied to find out how well they are able to detect zero-day network-based attacks. A 2014 paper tested zero-day threat detection on Snort, a network intrusion detection system. The Snort IDS was configured using an official rule set that was released on November 14\textsuperscript{th}, 2006 and was tested on 356 network attacks. A total of 183 of the tested attacks occurred after the rule set was published and thus were considered zero-day attacks by the IDS. The initial results from this study showed that Snort was able to detect 17\% of the tested zero-day threats but after considering false positives the authors posit a conservative zero-day detection rate is around 8\%~\cite{H_Holm}.

Initial research of network intrusion detection systems that involved machine learning methods brought to light two major shortcomings that will motivate the approach to solving this task. First, a majority of surveyed papers used a single dataset to train and test models while only holding out a subset of attacks from the train set for the test set. For example, in the DARPA 1998 dataset, the back attack (a DoS attack where HTTP requests are sent with URL's that contain many front slashes) is included in the train set and the Apache2 attack (a DoS attack where HTTP requests are sent that contain many headers) is included solely in the test set~\cite{DARPA98_ppt}. How different are these attacks and is this really considered zero-day detection? Second, the most used datasets are over twenty years old~\cite{ieee_survey}. Attack signatures have changed tremendously since these datasets were published, so how valid are these results in the present?

State of the art results in this area of research tend to skew heavily positive which introduces skepticism of their validity.  For example, N.B Amor et al. used a decision tree classifier and a naive Bayes classifier to detect DoS attacks in the KDD Cup 1999 dataset, a derivative of the DARPA 1998 dataset. Results show a 97\% DoS attack detection rate while keeping the false positive rate below 1\%~\cite{Amor}. This paper as well as many others that were surveyed in~\cite{ieee_survey} suffer from the same two flaws highlighted previously which the following work will attempt to solve. 

\section{Methods and Theory}

\subsection{Project Developments}

This project started with the idea of classifying network attacks using pcap raw network traffic data. Early in the project, it was difficult to find datasets that not only included the raw network traffic, but also included a mapping to label each packet as either benign or attack. Once three usable datasets were found, the focus of the project shifted to specifically classifying DDoS attacks since those were the attacks common to all datasets. Due to the size of the network traffic datasets offered and the computing power required to build the type of models proposed for this project, the scope of the project was set to smaller scenarios, like that of a small business. Many of the training data sets created in this project are in the range of 75 to 100k network packets long, which could be reasonable for monitoring a small business in real time. 

The original plan for testing these different datasets was to train on the oldest set, test on a newer set of attacks, and then perform the final model evaluations on the newest data set. Training solely on one older dataset with older benign and attack traffic proved difficult and did not yield good classification results when trying to classify newer attacks. This led to the creation of the different testing and training scenario datasets – five datasets in total. The first four of these scenario datasets were created to evaluate the best feature extraction technique and machine learning model architecture. These first four scenarios were created with a mixture of two of the datasets to see how much of the newer benign traffic effected the classification results. The final scenario was created for evaluating the best performing models chosen from the testing results using scenarios one through four.

Once the scenario datasets were created, the testing commenced to find the best feature extraction and machine learning model combination to use for the final evaluation. During testing, the testing plan required frequent adjustment. For example, there were certain feature and ML model combinations that did not make sense to use. Additionally, due to some poor results in the final evaluation, further testing was performed near the end of the project to look for insights into why some of these models did not perform as well as expected.

\subsection{Datasets}

This project includes the use of three different datasets: DARPA 1998~\cite{DARPA98}, SUEE 2017~\cite{SUEE}, and CIC DDoS 2019~\cite{CICDDoS2019}. These three datasets include different attack types and benign traffic patterns and were used throughout the project for creating feature extraction techniques and train/test datasets for machine learning classification models. The 1998 and 2017 datasets were used primarily for feature extraction and classification testing, whereas the 2019 dataset was used only for the final testing evaluations. Table 1 includes a summary of attack types and internet protocols included with each dataset.

The DARPA 1998 dataset was created by Lincoln Laboratory at MIT for the purpose of network intrusion detection research. MIT synthesized and recorded benign and malicious network traffic in a sandbox network environment. Nine weeks of network-based attacks were recorded. The data included both train and test datasets. ~\cite{DARPA98}. 

The 2017 SUEE dataset includes network traffic captured from Ulm University’s web server. The dataset includes multiple simulated DDoS attacks. The attack and benign data were captured over two sessions:  a twenty-four hour duration and a separate eight day duration. There is a total of three different attacks in this dataset that occur  150 different times throughout the eight day and twenty-four hour files. ~\cite{SUEE}. 

The CIC DDoS 2019 dataset was created by the Canadian Institute for Cybersecurity (CIC), located at the University of New Brunswick in 2019. This dataset was created to have a better engineered and more diverse set of attacks to be used for the purposes of  DDoS attack detection. The dataset includes a number of TCP and UPD based attacks, as well as synthetically generated benign traffic. The dataset includes raw pcap files for the purposes of feature extraction. ~\cite{CICDDoS2019}.

\begin{table}[h]
  \caption{Dataset Summary}
  \centering
  \begin{tabular}{ p{0.1\textwidth}p{0.1\textwidth}p{0.5\textwidth}} 
    \cmidrule(r){1-3}
    Dataset    & Year     & DDoS Attack Types \\
    \midrule
    DARPA & 1998 & back, land, neptune, pod, smurf, syslog, teardrop ~\cite{DARPA98} \\

    \midrule
    SUEE & 2017 & slowloris, slowhttp ~\cite{SUEE} \\

    \midrule
    CIC DDoS & 2019 & NTP, PortMap, DNS, NetBIOS, SNMP, LDAP, MSSQL, UDP, UDP-Lag, SYN ~\cite{CICDDoS2019}\\

    \bottomrule
  \end{tabular}
 \end{table}

\subsection{Project Approach - Testing Scenarios}

Using the three datasets from 1998~\cite{DARPA98}, 2017~\cite{SUEE}, and 2019~\cite{CICDDoS2019}, different test/train scenario datasets were constructed to explore the effect the benign traffic make-up had on the classification of new attack types. A total of five different test/train scenarios were created. The first four scenarios were used to evaluate the best model and feature extraction combination during testing. The last scenario uses the attacks from the 2019 dataset for the final model evaluation. A detailed summary of the test/train scenario dataset make-ups is provided in Table 2.

The different scenario datasets were constructed manually based on the type of benign activity and size of the attacks found in each respective dataset. Many of the attacks in the 1998 dataset were of varying length and quantity. When constructing the different scenario datasets, several different occurrences of the same attack were included. Some attacks that spanned over too many data packets when compared to the other attacks were shortened. Additionally, the benign traffic was selected so that there was a variety of different internet protocols present in the newly constructed dataset. Scenario 1 focuses on training entirely on DARPA 1998, whereas Scenario 2 introduces some of the 2017 benign traffic. One note is that the 2017 benign traffic being trained on is different from the traffic in Scenario 1-4’s testing set. Scenarios 3 and 4 explore the effect of taking out some of the 1998 data and replacing it with the 2017 benign traffic. Finally, the last scenario created utilizes the 2019 dataset for the final model evaluation.

\begin{table}[h]
  \caption{Project Train/Test Scenarios}
  \centering
  \begin{tabular}{ p{0.1\textwidth}p{0.35\textwidth}p{0.3\textwidth}p{0.1\textwidth}p{0.1\textwidth} } 
    \cmidrule(r){1-5}
    Scenario    & Training Set Description     & Testing Set Description & Training Count & Testing Count \\
    \midrule
    1 & DARPA 98 Attacks & SUEE 17 Attacks & 72k & 40k\\
     & DARPA 98 Benign (old protocols) & SUEE 17 Benign &  & \\
     & DARPA 98 Benign &  &  & \\
    \midrule
    2 & DARPA 98 Attacks & SUEE 17 Attacks & 97k & 40k\\
     & DARPA 98 Benign (old protocols) & SUEE 17 Benign &  & \\
     & DARPA 98 Benign &  &  & \\
     & SUEE 17 Benign &  &  & \\
    \midrule
    3 & DARPA 98 Attacks & SUEE 17 Attacks & 97k & 40k\\
     & DARPA 98 Benign &  SUEE 17 Benign &  &  \\
     & SUEE 17 Benign &  &  & \\
    \midrule
    4 & DARPA 98 Attacks & SUEE 17 Attacks & 100k & 40k\\
     & SUEE 17 Benign & SUEE 17 Benign &  & \\
    \midrule
    Final & DARPA 98 Attacks & CIC DDoS 19 Attacks & 23k & 43k\\
     Holdout & SUEE 17 Benign & CIC DDoS 19 Benign &  & \\
     Eval & SUEE 17 Attacks &  &  & \\
     & CIC DDoS Benign &  &  & \\
    \bottomrule
  \end{tabular}
 
\end{table}

\subsection{Project Approach - Feature Extraction}

\subsubsection{Manual Feature Extraction}

Manual feature extraction allowed for the application of knowledge of network traffic and DDoS attacks to extract meaningful information from the datasets. This is also a useful skill to have when facing real world datasets, as it is important to understand the type of data being classified. Additionally, another advantage that manual feature extraction can have over an automated method is the ability to explain the output to other researchers. This could also lead to a better understanding of the dataset if certain features perform better when being used for classification purposes.

The manual features for this project were extracted from the “info” field that is provided by common network data packet analyzers. The information from this specific field was extracted via Wireshark and then was imported to a python script. The manual features extracted by the python script are mostly keywords, phrases, or internet protocol ID numbers. A total of sixteen manual features were extracted per data packet.

The second set of manual features were created based on the first set of sixteen features, but were also batched in groups of ten packets based on the source IP address. The rationale behind the creation of this second set of manual features is that DDoS attacks tend to happen in groups, so batching of the manual features in groups may prove to be useful. The newly batched features included counts of the original sixteen manual features (i.e., how many times did feature one show up across the last ten packets from this IP address, how many time did feature two show up across the last ten packets, etc.). In addition to counts for the original sixteen features, additional features were created that were specific to the batch process. For example, one of the batching specific features is looking for how many unique protocol ID numbers occurred in the last ten data packets. The batched set of manual features included a total of twenty-two features.

\subsubsection{NLP Feature Extraction}

Natural language processing (NLP) was used in this project as an automated feature extractor. Similar to the manual feature extraction process, the NLP features were extracted from the “info” field that is provide by Wireshark. The strings provided in this info field for each packet were tokenized and then ran through a NLP transformer model. The NLP transformer model outputs 768 sentence embeddings for each network packet processed through the model~\cite{Distilbert}. The 768 sentence embedding features per network packet were then used as features for classification.

The NLP transformer model used for feature extraction is the Hugging Face transformer, Distilbert~\cite{Distilbert}. The Distilbert transformer model is based off Google’s BERT model~\cite{BERT}, however Distilbert has less model parameters compared to BERT~\cite{Distilbert}. Since the model has less parameters, the time to run the model and to fine tune the model is reduced significantly~\cite{Distilbert}. Less parameters also means less performance compared to BERT, however the reduction in performance is small~\cite{Distilbert}. The Distilbert model was fine-tuned using the 1998 attack traffic, the 1998 benign traffic, and the 2017 benign traffic.

\subsubsection{Autoencoder Feature Extraction}

Due to its natural ability in learning compressed representations of input data, the autoencoder (AE) model architecture was employed as the second automated feature extraction technique studied during this project. The feature extraction AE created for this research utilizes a network intrusion dataset's raw pcap data in the form of a Wireshark hex dump. Each network packet is represented as a hexadecimal byte that is first padded to a fixed length of 1,518 bytes, the maximum Ethernet frame size, before being converted and parsed into 12,144 binary bits for handling by the AE. 

The AE model compresses the 12,144 bit input down to 128 bits at the encoder output layer to be used as the features for classification and the structure of the model can be seen in Equations (1)-(4).

\begin{equation}
    \mathbf{\emph{H}}_{1} = f_{1}(\mathbf{\emph{X}}) = (\mathbf{\emph{W}}_{1} \mathbf{\emph{X}} + \mathbf{\emph{b}}_{1}) \textrm{ (encoder hidden layer)}
\end{equation}

\begin{equation}
    \mathbf{\emph{H}}_{2} = f_{2}(\mathbf{\emph{H}}_{1}) = \sigma(\mathbf{\emph{W}}_{2} \mathbf{\emph{H}}_{1} + \mathbf{\emph{b}}_{2}) \textrm{ (encoder output layer)}
\end{equation}

\begin{equation}
    \mathbf{\emph{H}}_{3} = f_{3}(\mathbf{\emph{H}}_{2}) = (\mathbf{\emph{W}}_{3} \mathbf{\emph{H}}_{2} + \mathbf{\emph{b}}_{3}) \textrm{ (decoder hidden layer)}
\end{equation}

\begin{equation}
    \hat{\mathbf{\emph{X}}} = f_{4}(\mathbf{\emph{H}}_{3}) = \sigma(\mathbf{\emph{W}}_{4} \mathbf{\emph{H}}_{3} + \mathbf{\emph{b}}_{4}) \textrm{ (decoder output layer)}
\end{equation}

A sparsity constraint is imposed on the encoder output layer to mimic the natural sparsity of the raw network packet input. A mean squared error loss function is used along with a KL divergence regularization parameter to measure the overall reconstruction loss. The combined loss function is represented in Equation (5). The AE was trained using DARPA 1998 and SUEE 2017 benign data only to achieve an anomaly-based feature extraction method. 

\begin{equation}
    \mathcal{L} = {1\over2N} \sum_{i=1}^{N} ||\mathbf{\emph{x}}_{i} - \hat{\mathbf{\emph{x}}}_{i}||_{2}^{2} + \alpha \sum_{i=1}^{N} \mathbf{\emph{x}}_{i} \log{\mathbf{\emph{x}}_{i}\over\hat{\mathbf{\emph{x}}}_{i}} + \beta \sum_{i=1}^{m} |\mathbf{\emph{H}}_{2}^{i}|
\end{equation}

\subsubsection{CICFlowMeter Feature Extraction}

The CICFlowMeter tool was developed by researchers at the University of New Brunswick’s Canadian Institute for Cybersecurity and is a popular method for creating network features to be used in network intrusion detection. The tool reads in a pcap file and creates network flows along with seventy-six statistical features per flow. A network flow is defined as a sequence of packets that are sent between a unique source machine and a unique destination machine. A small sample of features that are created for each flow include flow duration, total fwd/bwd packets, flow packets per sec, average packet length, and packet flags. A full list of features that are created by the CICFlowMeter tool can be found in~\cite{CICFlowMeter}. The CICFlowMeter tool will be used as a baseline feature extraction method that will help determine how well this project's novel feature extraction methods perform.  

\subsection{Project Approach - Classification Models}

\subsubsection{MLP}

A survey of existing literature in this field showed that a respectable number of researchers found success using basic machine learning techniques, such as support vector machines, clustering methods, and simple artificial neural networks~\cite{ieee_survey}. As a result, it was important to test this project's feature extraction algorithms using a "vanilla" neural network in order to really understand how well each method works without the aid of a more complicated network. To achieve this a shallow feedforward multilyaer perceptron model was developed. 

The feedforward multilayer perceptron (MLP) neural network model used during testing had three dense layers and three dropout layers to prevent over-fitting. Each hidden dense layer uses a ReLu activation function, and the last output layer uses sigmoid activation. The size of the dense layers were 0.75x the size of the input layer. The loss function used for this model was binary cross-entropy, which was appropriate since our classification models are performing binary classification: 0 for benign, 1 for attack.

A few different variations of this MLP model were tested in the results section. The first variation includes not having drop layers between the dense layers of the model. The other variations of this model include the use of K-means clustering. First, K-means clustering was ran on each scenario to give a benchmark performance of how well the packets being tested fit into two clusters. A second test included the combination of K-means clustering and the MLP output, where the output of the two models were averaged, then assigned a classification. The third use of K-means clustering included using the output of K-means clustering as an additional feature for the MLP to classify.

\subsubsection{LSTM}

The second model architecture that was developed for testing the three new feature extraction algorithms introduced previously was a long short-term memory (LSTM) network. LSTM models are known for their ability to handle sequential data which lends itself well to this task as DDoS attack signatures and network traffic in general are sequential in nature. The math behind a single LSTM cell can be seen in Equations (6)-(11).

\begin{equation}
    \mathbf{\emph{c}}_{t} = f_{t}(\mathbf{\emph{h}}_{t-1}, \mathbf{\emph{x}}_{t}) \odot \mathbf{\emph{c}}_{t-1} + i_{t}(\mathbf{\emph{h}}_{t-1}, \mathbf{\emph{x}}_{t}) \odot a_{t}(\mathbf{\emph{h}}_{t-1}, \mathbf{\emph{x}}_{t}) \textrm{ (cell state vector)}
\end{equation}

\begin{equation}
    \mathbf{\emph{h}}_{t} = o_{t}(\mathbf{\emph{h}}_{t-1}, \mathbf{\emph{x}}_{t}) \odot \tanh(\mathbf{\emph{c}}_{t}) \textrm{ (hidden state vector)}
\end{equation}

\begin{equation}
    f_{t}(\mathbf{\emph{h}}_{t-1}, \mathbf{\emph{x}}_{t}) = \sigma(\mathbf{\emph{W}}_{f} \mathbf{\emph{x}}_{t} + \mathbf{\emph{U}}_{f} \mathbf{\emph{h}}_{t-1} + \mathbf{\emph{b}}_{f}) \textrm{ (forget gate activation vector)}
\end{equation}

\begin{equation}
    i_{t}(\mathbf{\emph{h}}_{t-1}, \mathbf{\emph{x}}_{t}) = \sigma(\mathbf{\emph{W}}_{i} \mathbf{\emph{x}}_{t} + \mathbf{\emph{U}}_{i} \mathbf{\emph{h}}_{t-1} + \mathbf{\emph{b}}_{i}) \textrm{ (input/update gate activation vector)}
\end{equation}

\begin{equation}
    o_{t}(\mathbf{\emph{h}}_{t-1}, \mathbf{\emph{x}}_{t}) = \sigma(\mathbf{\emph{W}}_{o} \mathbf{\emph{x}}_{t} + \mathbf{\emph{U}}_{o} \mathbf{\emph{h}}_{t-1} + \mathbf{\emph{b}}_{o}) \textrm{ (output gate activation vector)}
\end{equation}

\begin{equation}
    a_{t}(\mathbf{\emph{h}}_{t-1}, \mathbf{\emph{x}}_{t}) = \tanh(\mathbf{\emph{W}}_{a} \mathbf{\emph{x}}_{t} + \mathbf{\emph{U}}_{a} \mathbf{\emph{h}}_{t-1} + \mathbf{\emph{b}}_{a}) \textrm{ (cell input activation vector)}
\end{equation}

The model consists of four LSTM layers each with 128 hidden units, dropout layers after each LSTM layer, and a dense layer with a sigmoid activation function for the final prediction. Similar to the MLP model, the dropout layers were added to prevent overfitting and a binary cross-entropy loss function was used.

\subsection{Testing Approach}

This project was organized into three major stages, each of which required a different set of testing and milestones to complete. The first major stage in this project was developing feature extraction techniques. The manual, NLP, and autoencoder feature extraction techniques were developed and tested on smaller datasets at the start. This was to give an indication of how well the features were describing the network data. Note that these smaller datasets were different from the scenario datasets mentioned earlier. Figure 6 in the Appendix shows how the performance of the feature extraction techniques were tracked via Excel spreadsheets.

With the feature extraction techniques finalized, stage two of the project included testing a combination of machine learning models, test/train scenarios one to four, and the different feature extraction methods. Figures 7 – 13 in the Appendix show the results from the second stage of testing. All results were recorded via an Excel spreadsheet, and confusion matrices with conditional formatting were created for each test. 

Finally, with the results from stage two, stage three included using the best model and feature extraction combinations to classify the last dataset held out for final evaluation. Further additional tests were also performed at this time, including performing a baseline test with the CICFlowMeter results, as well as comparing the timing of both models. One note is that the testing results reported in this project is on a per packet basis, not a per attack basis. The metrics reported note how many individual attack packets were classified correctly.

\section{Results}

\subsection{MLP Results}

The results of the MLP model variations are found in Figures 7 -10 of the Appendix. Figures 7 and 8 show that the unbatched version of the manual features performed better than the batched version, however, the results across the manual features are generally poor and inconsistent. The reason for this could have been that the manual feature extraction method does not pull as many features as the other compared models, and batching the data causes there to be too few samples and features to train the model on. K-means clustering helped increase the classification results of the attack traffic using the manual features, but also decreased the classification rate of the benign traffic. Figures 9 and 8 in the Appendix show the results using the NLP features. The NLP performs much better than the manual features, especially the features that were batched. The batched NLP feature results are much more consistent across all tests. The best performing model in this set was the MLP with dropout layers on scenario 3 using NLP batched features. This model was selected to be used for final evaluation. 

One note is that CICFlowMeter and AE feature extraction techniques did not yield any attack classifications. In the case of the AE features, these features may have required a more time series approach, which is why the LSTM seems to perform better with this feature set. In the case of the CICFlowMeter, it is possible that the scenario datasets created for the purposes of testing contain too many sliced together packets. The CICFlowMeter relies on a flow from destination to source address. The datasets created for the scenario testing may need to be segmented for the CICFlowMeter tool to work.

\subsection{LSTM Results}

The LSTM model was tested on the three feature extraction techniques developed for this project but it was not tested on the CICFlowMeter baseline feature extraction method. The reason behind this was that by the creation of CICFlowMeter's network flows, the time series nature of the raw network data was removed. Three different input time-step variations were tested for each dataset scenario and each feature extraction algorithm. Results from all of the following tests can be seen in Figures 11-13 located in the Appendix. 

First a set of tests were run where the input to the model was only the current network packet, i.e., a time-step of one. Overall, the classification results for this input set up were quite poor, with the best results, 49\% attack classification rate, coming from the LSTM model trained on scenario three data while using the NLP generated features. These findings were no surprise as majority of DoS attacks occur over multiple network packets which makes it difficult for a LSTM model with a time-step of one to detect. 

To account for the time-series nature of DoS attack signatures, the time-step was increased to contain the last sixty-four (sixteen for the NLP features) packets in sequential order. All tests were re-run with the increased time-step and although the manual and AE features still produced poor results, the NLP features saw a decent jump in performance. The LSTM model trained with NLP features on the scenario three dataset once again achieved the highest classification rate at 61\% DoS attack detection. 

Finally, the input time-step was held constant but instead of containing the last sixty-four (sixteen for the NLP features) sequential packets, the inputs were now grouped by flow. The current packet determined the source and destination machines and the past sixty-four (sixteen for the NLP features) packets sent between the two were included in the input. If there were not enough packets to completely fill the input, it was padded with zero vectors until full. Unlike the creation of network flows with the CICFlowMeter tool this process did in fact preserve the time-series nature of the network data by only using packets that have already been sent. After re-running all tests a final time, the results showed minor improvements in attack classification and false positives using the AE and NLP features. 

Similar to the MLP model results, the best performing LSTM model was chosen to be the one trained on the scenario three dataset, using the NLP generated features, and an input time-step of last sixteen packets grouped by flows. This model was picked to be used for final evaluation since it had zero false positives and one of the highest DoS attack detection rates of 61.32\%.

\subsection{Feature Extraction Baseline Comparison}

To further test the two novel automated feature extraction algorithms developed for this project against the CICFlowMeter baseline a simple 80/20 train/test split was performed for each dataset scenario. In order to fully capture how well each feature extraction method compared to the CICFlowMeter baseline the MLP classification model was used. The results for this set of tests can be seen in Figure 14 located in the Appendix. 

The NLP generated features were seen to produce the best attack detection on scenarios one to three while performing poorly on scenario four. The AE feature extraction method returned the opposite results, achieving the best attack classification on scenario four while performing poorly on scenarios one to three. These findings are consistent with the initial MLP and LSTM testing results discussed previously and further motivates the choice to use the NLP generated features for the final evaluation. 

The CICFlowMeter feature extraction baseline returned consistent results across the board but achieved the best attack classification on scenarios three and four which were the most realistic/real world datasets. Overall, the NLP generated features performed on par with the CICFlowMeter, even outperforming it on scenarios one, two, and three. Once again this helps solidify the choice of using the NLP features for final evaluation.

\subsection{Attack Timing and Model Comparison}

The MLP and LSTM machine learning models were compared on timing and classification performance using NLP features and the scenario three dataset. In regard to timing, it was noted that both the MLP and LSTM models would lag an average of a thousand packets before one of the first attacks in the test set was classified. However, once the attacks started happening more frequently in the test set, the timing improved for both models.  Figures 20 and 21 in the Appendix show two examples of timing, one for a shorter attack and the other for a longer attack. The performance of the LSTM and MLP are similar in terms of picking up the attack, however the overall performance for both seems to drop when the attacks become longer.

Figures 15 to 19 in the Appendix show a side-by-side comparison of the MLP and LSTM in terms of attack classification across the different IP addresses in the test set. The red highlighting in the table notes if a large portion of the attack packets were not classified by each model. The yellow highlighting in the table notes if a large portion of benign packets were classified as attack packets by the models. In general, the MLP had a harder time with classifying the benign packets, as there are a few IP address where the MLP thinks there is an attack when there is not. In terms of missing attacks, the MLP and LSTM seemed to have very similar performance – if one model is in the red, it is likely the other model is as well. Figures 17 and 18 show the most amount of missed packets. The IP addresses noted in these two figures correspond to a slowHTTP attack~\cite{SUEE}, which looks to be the most difficult for the models to classify.

\subsection{Final Evaluation}

The best MLP and LSTM models were chosen from the previous batch of testing to further test their capabilities at detecting zero-day network threats. This was completed by re-training each model and testing it against the 2019 CIC DDoS dataset which contained even more recent DDoS attacks. This set of tests also simulated a real-world scenario where an initial model was trained to detect zero-day attacks, a novel zero-day attack was carried out, and then the initial model was re-trained/fine-tuned on this new attack. Final evaluation results can be seen in Figure 22 located in the Appendix.

The MLP model was fully retrained using scenario three data along with the original SUEE test set and a subset of benign traffic from the CIC DDoS dataset. Both unbatched and batch NLP features were used to retrain the model. Testing on the new 2019 dataset showed a significant drop in performance with the unbatched NLP features catching only around 35\% of the new DoS attacks. 

The LSTM model was first tested on the new dataset without any re-training or fine-tuning and was able to detect 31\% of the novel attacks. The model was then re-trained using scenario three data along with the original SUEE test set and a subset of benign traffic from the CIC DDoS dataset. Testing on the re-trained LSTM model showed a slight performance increase with the model now being able to detect 33\% of the attacks while also lowering the false detection rate by 0.5\%.

An 80/20 train/test split was performed on the 2019 evaluation dataset to baseline the NLP feature extraction algorithm one last time. Results confirmed that the NLP features were still detecting network attacks at a slightly better rate than the CICFlowMeter baseline. 

On average around a 30\% decrease in zero-day attack detection was seen between the initial testing results and the final evaluation results. The significant drop in performance may be attributed to how the 2019 dataset was constructed vs the datasets that were tested on in scenarios one through four. The 2019 dataset included many pcap files where the activity within the file was identical. For example, a few hundred pcap files, each containing about 400k individual packets, would have almost identical activity. To create a more diverse set of training and testing files using the 2019 dataset, the test and train files were created using smaller slices of several different pcap files. This method of assembling a test set for the 2019 evaluation may attribute to some of the poor performance seen. In comparison, the test set for scenarios one to four included just one continuous portion of the data provided in the dataset. The decrease in the attack detection on the 2019 dataset could also be attributed to the benign traffic protocols. Not only are the attacks different in the 2019 dataset when compared to the 1998 and 2017 training sets, but the benign traffic has also changed. In the 2019 dataset, there are more protocols relating to the streaming of data, such as audio, videos, or gaming. This shift in normal internet activity may have also played a large role in the decrease of attack detection on the final evaluation dataset.

\section{Discussion and Conclusion}

\subsection{Small Business Recommendation}

After analyzing the findings from the numerous feature extraction and classification tests preformed during this project, three main small business recommendations were curated for the detection of zero-day DDoS attacks using machine learning techniques.

First, classification model training data must include benign traffic from the business's own network. This fact was made clear by comparing the poor test results that used scenario one for training, which included no benign traffic from the same distribution as the test set, and the much better test results that used scenarios two, three, or four for training, which did include benign traffic from the same distribution as the test set. By comparing test results from scenario two and scenario three or four, it can be posited that out of distribution benign traffic may still be included in the training data with limited to no negative impact as along as in distribution benign traffic is in the majority. 

Second, the NLP feature extraction algorithm achieved the best and most consistent zero-day DDoS attack detection across the board. Both the MLP and LSTM models trained using the NLP features vastly outperformed both the manual and AE features during the initial testing phase. The NLP generated features even performed better at detecting these zero-day threats than the CICFlowMeter baseline feature extraction method. 

Third, the classification model choice is not as important as the feature extraction process. By analyzing the attack timing results described in Section 4.4, one can see that both the MLP and LSTM models, in general, produce similar results in detecting attacks. Where the models differ slightly is in their false alarm rate, the MLP model had a more difficult time classifying benign packets where as the LSTM model really excelled at classifying benign packets.  

\subsection{Future Work}

Future work for this project beyond the scope of this course will include testing the attacks contained within the existing three datasets with benign traffic that is collected from home internet activity. As noted in the final recommendation, having a large, diverse set of benign traffic is critical for achieving an acceptable classification rate using the feature extraction techniques and classification models tested in this project. The home internet traffic will be collected using an NTAP device, or network tap. The network tap will be positioned between the home gateway and router, will be completely hidden from other network devices in the home, and will collect to and from benign data packets. The network tap device will export a tcpdump file to an external terminal or file location, where the data will be collected and used for creating new datasets. 

When collecting home internet traffic, a diverse set of internet activity will be used. Three main protocols will be utilized when collecting traffic: SFTP, TCP, and UDP. The SFTP data will be collected by performing a large SFTP file transfer. The TCP data will be collected by performing common internet tasks, such as send various emails with attachments. Finally, UDP data will be collected by streaming data, such as video or audio. Once these types of protocols are collected from the home internet traffic, training datasets will be constructed to make sure all three protocols are present throughout. Additionally, this benign traffic will be mixed with 1998 traffic, 1998 attacks, 2017 traffic, and 2017 attacks. This newly constructed dataset will then be used to classify the 2019 attacks to see if there is any improvement in classification accuracy.

The detection of zero-day network attacks may be viewed as a "cat and mouse game" between the attacker and intrusion detection system where the attacker is always trying to find ways to trick the intrusion detection system into misclassifying the attack. Adversarial machine learning attacks involve manipulating input data, for example by adding random noise to the input, before feeding it into a model for inference.  Due to the nature of this "cat and mouse game" and the methodology behind adversarial machine learning attacks, adversarial model training may prove useful in increasing zero-day threat detection rates.   

As a result of the sub-optimal final evaluation results for both the MLP and LSTM models, ensemble learning methods, such as stack generalization, may help increase the zero-day threat detection rate. Initial testing of the MLP model already employed a variation of stack generalization to some success, by averaging the output of a K-means clustering algorithm with the output of the MLP model. Since the LSTM model proved to be more acute at benign packet classification and the MLP model achieved better attack packet classification, stacking both models may help increase the overall results.   

\subsection{Challenges and Final Thoughts}

One major challenge of this project is working with real world datasets, specifically datasets that involve raw network traffic. Network traffic can often be unintuitive, and it may not be clear how to extract useful information. Datasets that provide raw pcap or tcpdump files also tend to have a very large ratio of benign traffic to attack traffic, making it difficult to create train and test datasets that have an appropriate amount of attack data. Another major hurdle in this project is that datasets that include raw network traffic typically label the packets as benign/attack by using an existing feature extraction technique. This often leads to a disconnect between the labels provided by the dataset and the raw network traffic, as the labeling is not one-to-one when compared to the raw data packets. This required manually labeling the raw network packets given the labels and new features provided, which introduces another possibility for error in the project.

Not only was it difficult to decipher these datasets, once they were in a usable state, the disc and memory space needed to process these datasets proved challenging to obtain. The datasets used for this project were hundreds of gigabytes in size. Due to their size and the sheer volume of packets contained in them, in the range of millions to tens of millions of packets, these datasets were also tricky to operate on in a timely manner. This fact was the driving force behind focusing this project on a small business solution, were the volume of packets contained in each dataset ranged from tens of thousands to hundreds of thousands. 

Even though zero-day attack detection has been studied extensively over the past twenty years, there are still a number of attacks that go undetected every year. As DDoS attack signatures evolve and become even more difficult to detect, machine learning techniques are starting to provide some much needed relief. The research completed during this project highlights that by combining automatic feature extraction algorithms with different classification model architectures real world zero-day DDoS attack detection is possible and that there is much potential to increase detection rates by deploying the methods discussed in Section 5.2.

\pagebreak

\pagebreak

{\bibliography{refs}}
\bibliographystyle{plainnat}

\pagebreak

\section*{Appendix - Dataset Information}

\begin{figure}[H]
  \centering
  \includegraphics[scale=0.45]{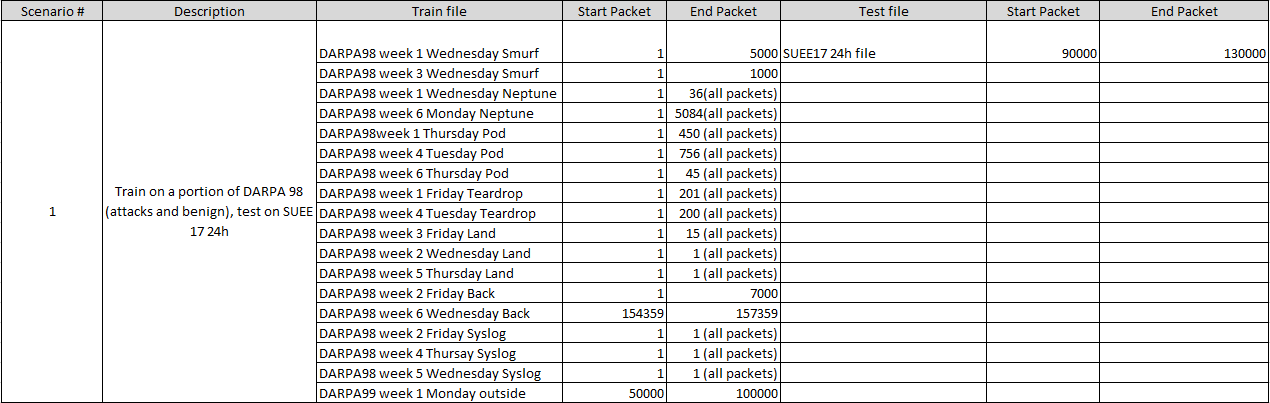}
  \caption{Scenario 1 Train/Test dataset packet makeup}
\end{figure}

\begin{figure}[H]
  \centering
  \includegraphics[scale=0.45]{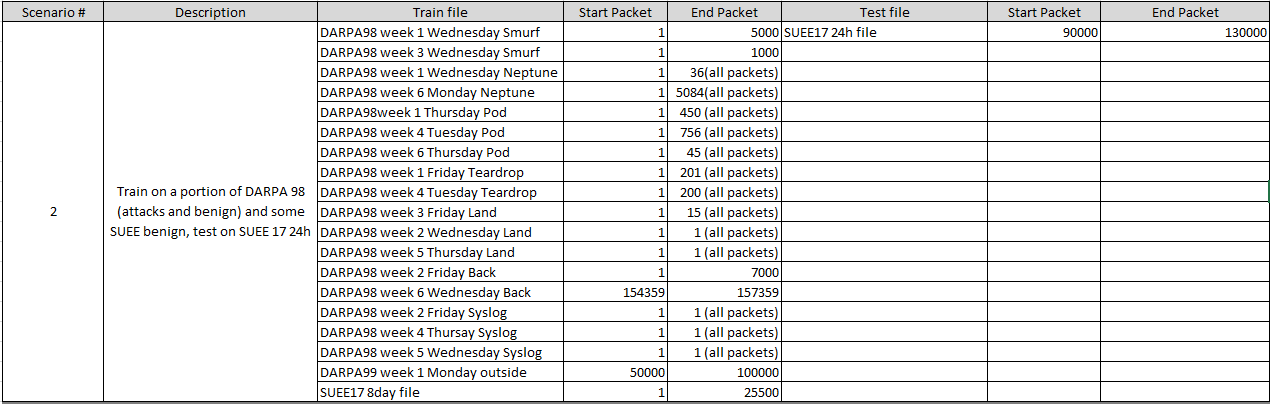}
  \caption{Scenario 2 Train/Test dataset packet makeup}
\end{figure}

\begin{figure}[H]
  \centering
  \includegraphics[scale=0.45]{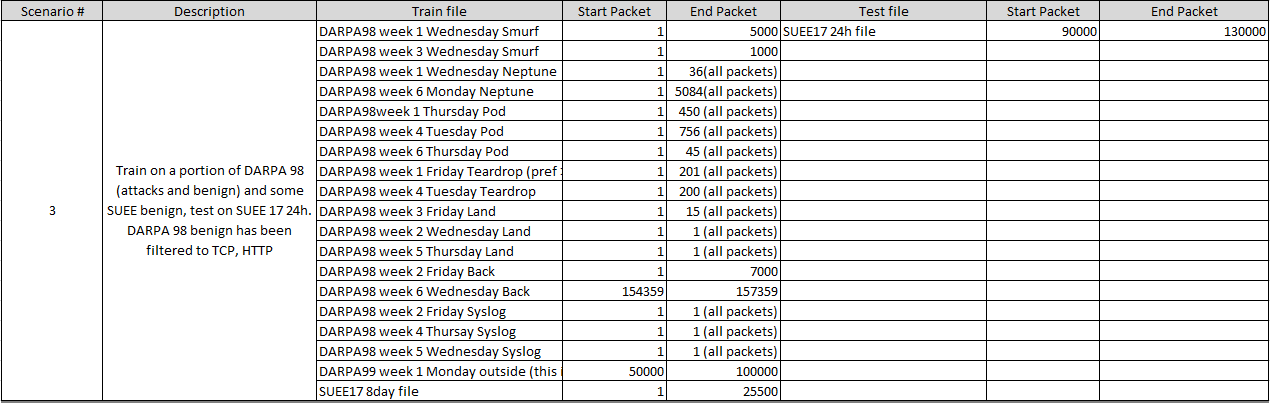}
  \caption{Scenario 3 Train/Test dataset packet makeup}
\end{figure}

\begin{figure}[H]
  \centering
  \includegraphics[scale=0.45]{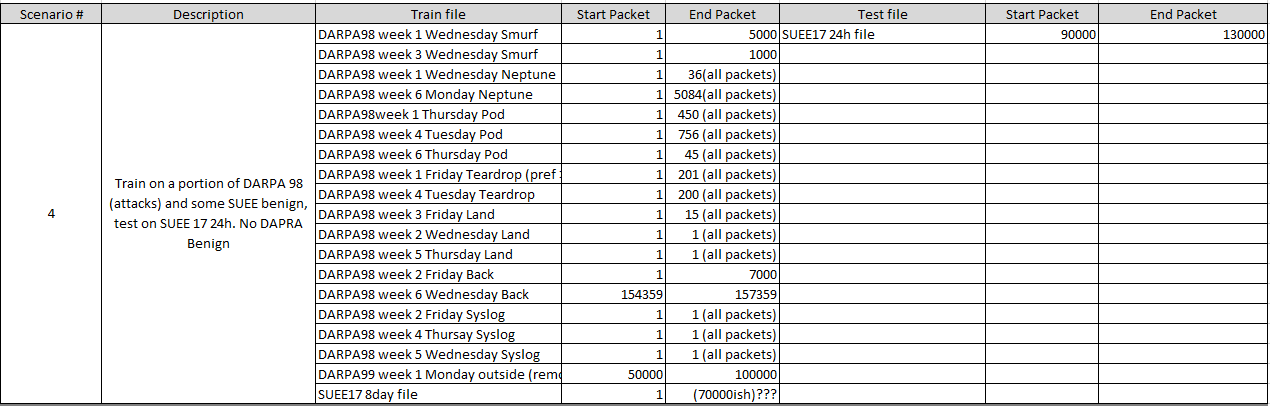}
  \caption{Scenario 4 Train/Test dataset packet makeup}
\end{figure}

\begin{figure}[H]
  \centering
  \includegraphics[scale=0.45]{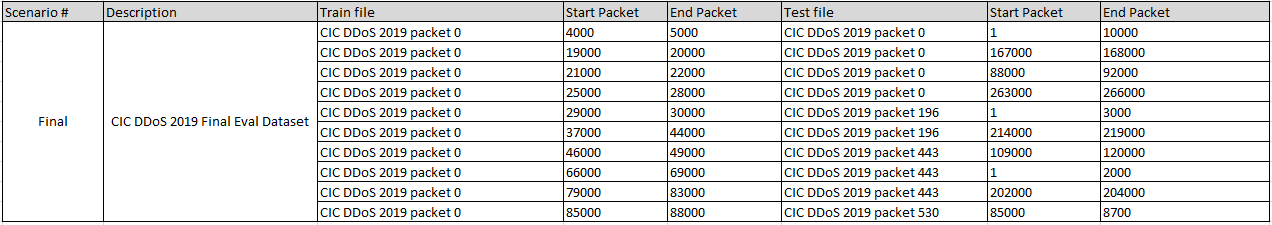}
  \caption{Final Evaluation Train/Test dataset packet makeup}
\end{figure}

\pagebreak

\section*{Appendix - Results}

\begin{figure}[H]
  \centering
  \includegraphics[scale=0.4]{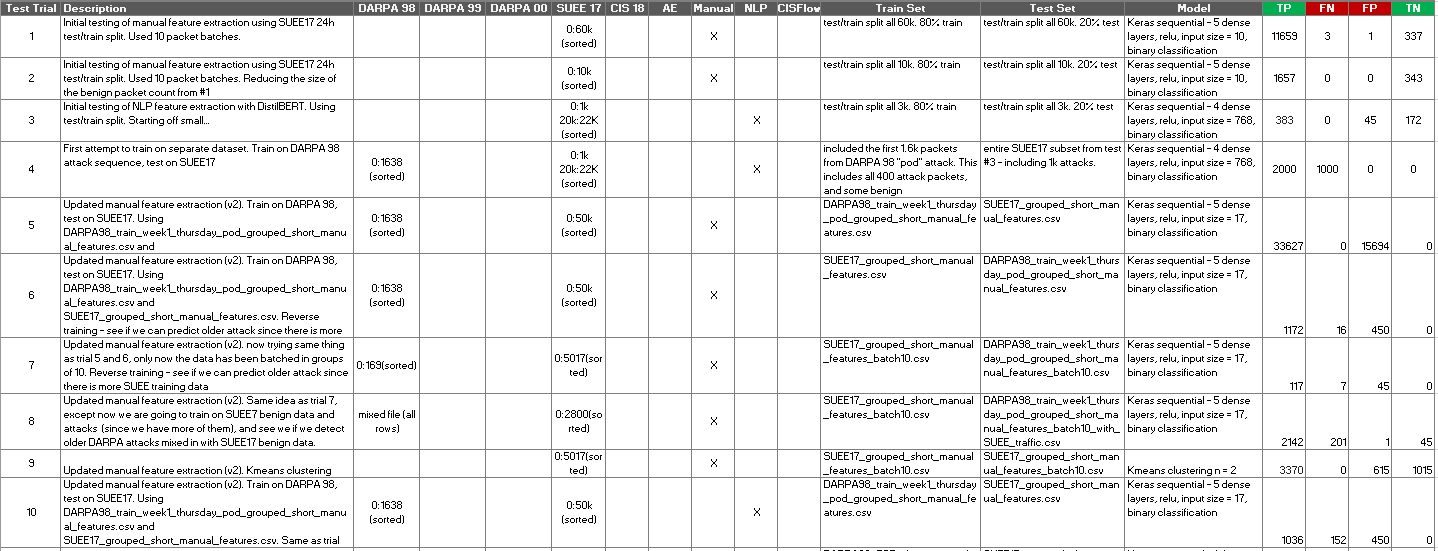}
  \caption{Feature Extraction Testing Results. Note this is a small sample of the test performed and tracked to show our tracking methods to determine the final feature extraction methods. }
\end{figure}

\begin{figure}[H]
  \centering
  \includegraphics[scale=0.8]{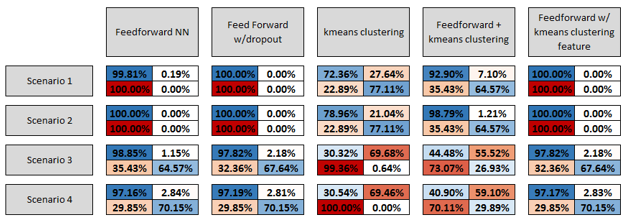}
  \caption{Feed-forward Neural Network Manual Feature Test Results}
\end{figure}

\begin{figure}[H]
  \centering
  \includegraphics[scale=0.8]{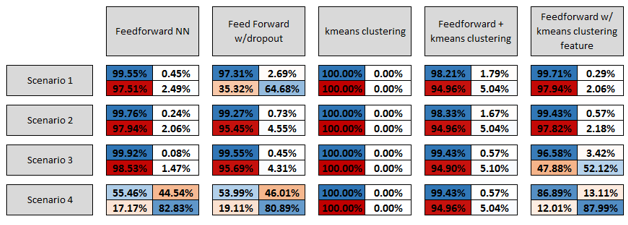}
  \caption{Feed-forward Neural Network Manual Batched Feature Test Results}
\end{figure}

\begin{figure}[H]
  \centering
  \includegraphics[scale=0.8]{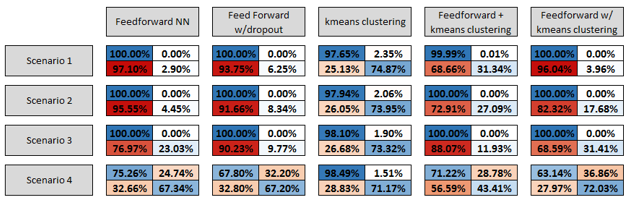}
  \caption{Feed-forward Neural Network NLP Feature Test Results}
\end{figure}

\begin{figure}[H]
  \centering
  \includegraphics[scale=0.8]{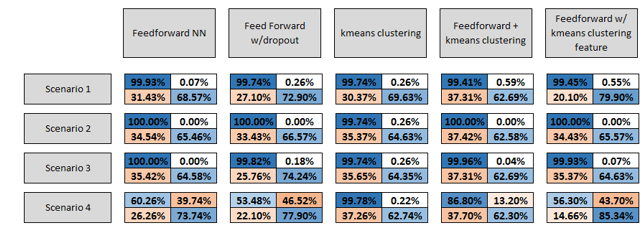}
  \caption{Feed-forward Neural Network NLP Batched Feature Test Results}
\end{figure}

\begin{figure}[H]
  \centering
  \includegraphics[scale=0.8]{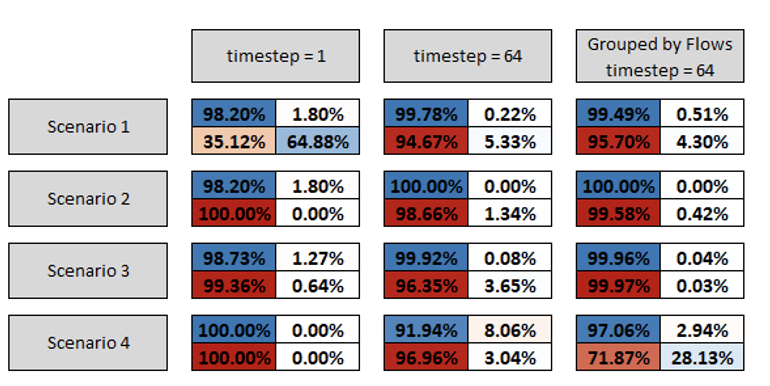}
  \caption{LSTM Manual Feature Results}
\end{figure}

\begin{figure}[H]
  \centering
  \includegraphics[scale=0.8]{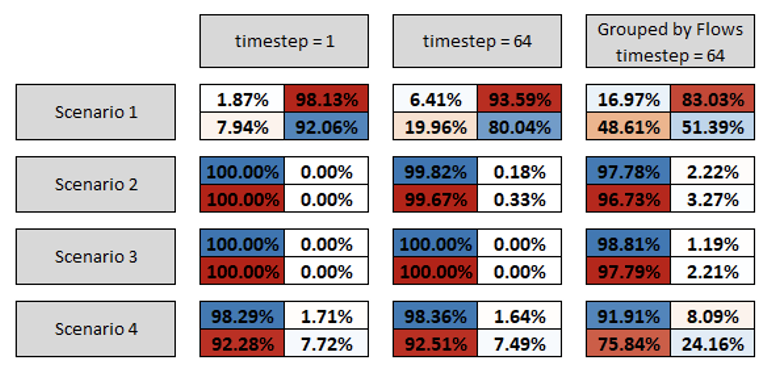}
  \caption{LSTM AE Feature Results}
\end{figure}

\begin{figure}[H]
  \centering
  \includegraphics[scale=0.8]{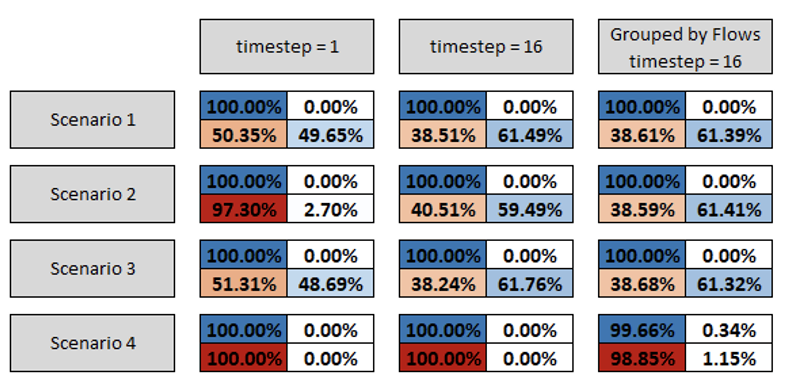}
  \caption{LSTM NLP Feature Results}
\end{figure}

\begin{figure}[H]
  \centering
  \includegraphics[scale=0.4]{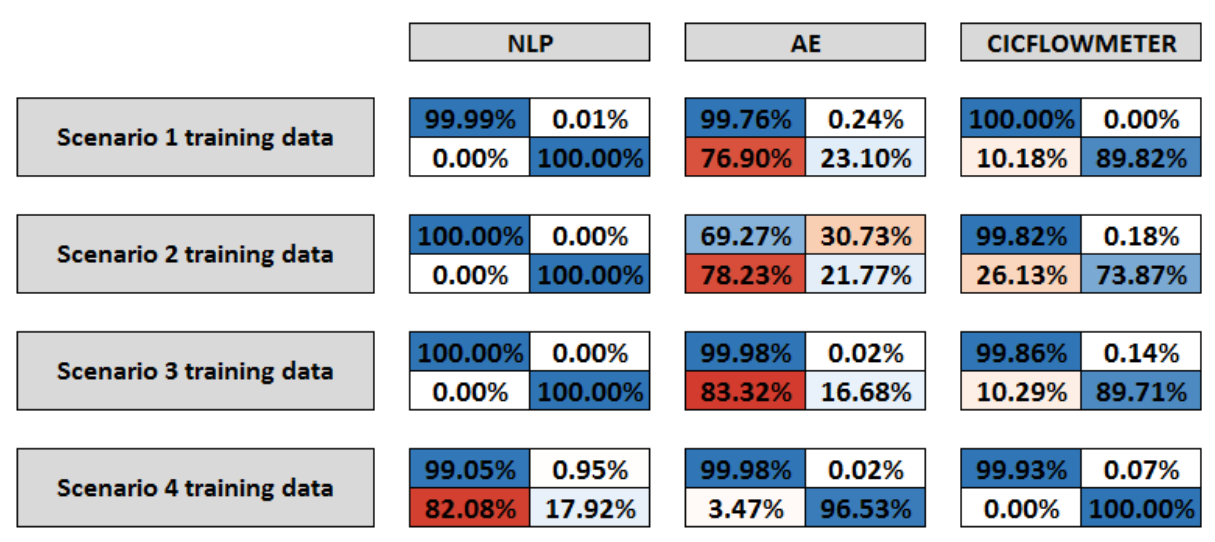}
  \caption{Feature extraction baseline comparison using an 80/20 train/test split on each dataset scenario.}
\end{figure}

\begin{figure}[H]
  \centering
  \includegraphics[scale=0.6]{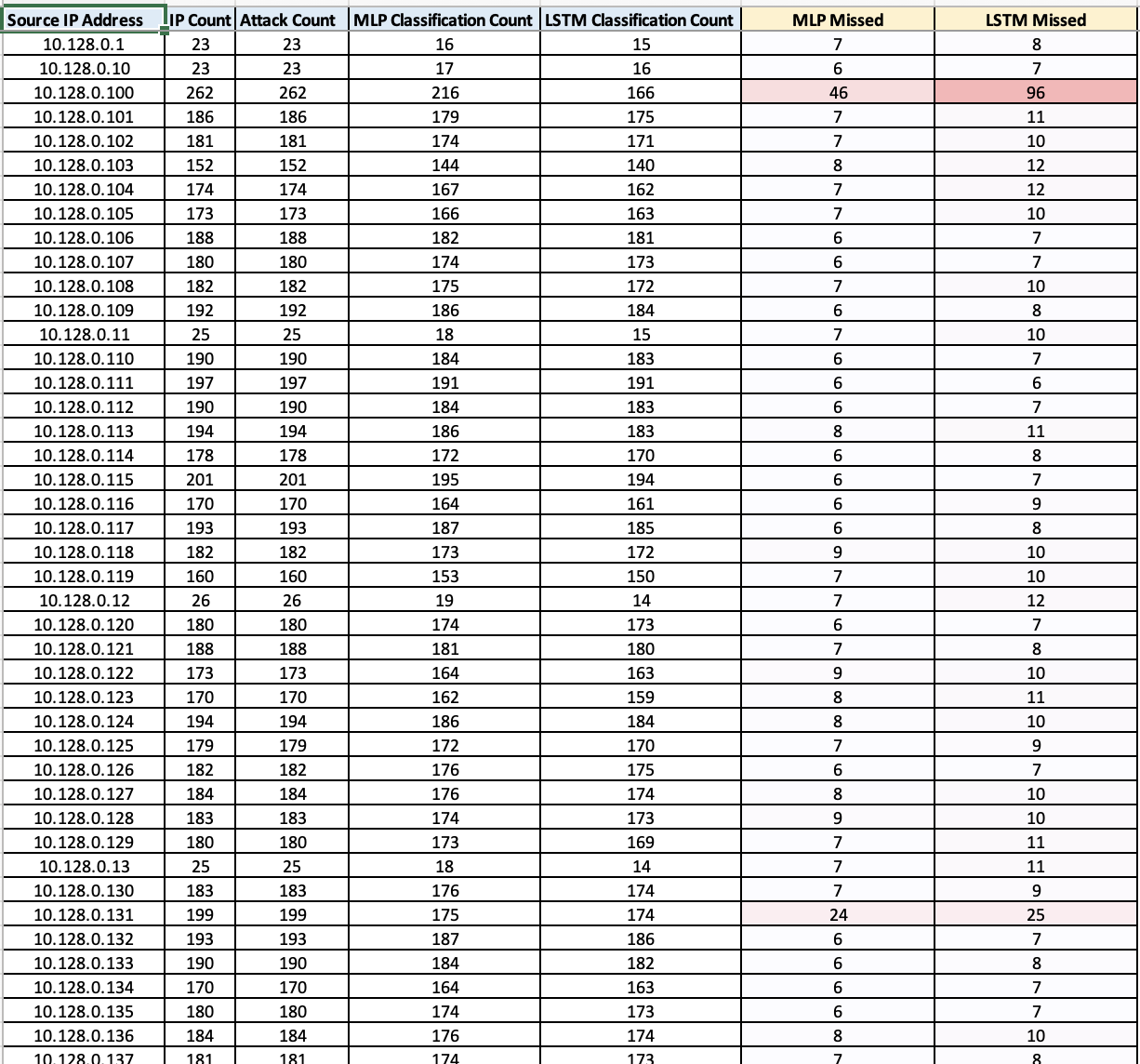}
  \caption{MLP and LSTM Model comparison. Red scale notes how many attack packets were missed by each mode. Yellow scale notes how many true benign packets were classified as attacks by the models.}
\end{figure}

\begin{figure}[H]
  \centering
  \includegraphics[scale=0.6]{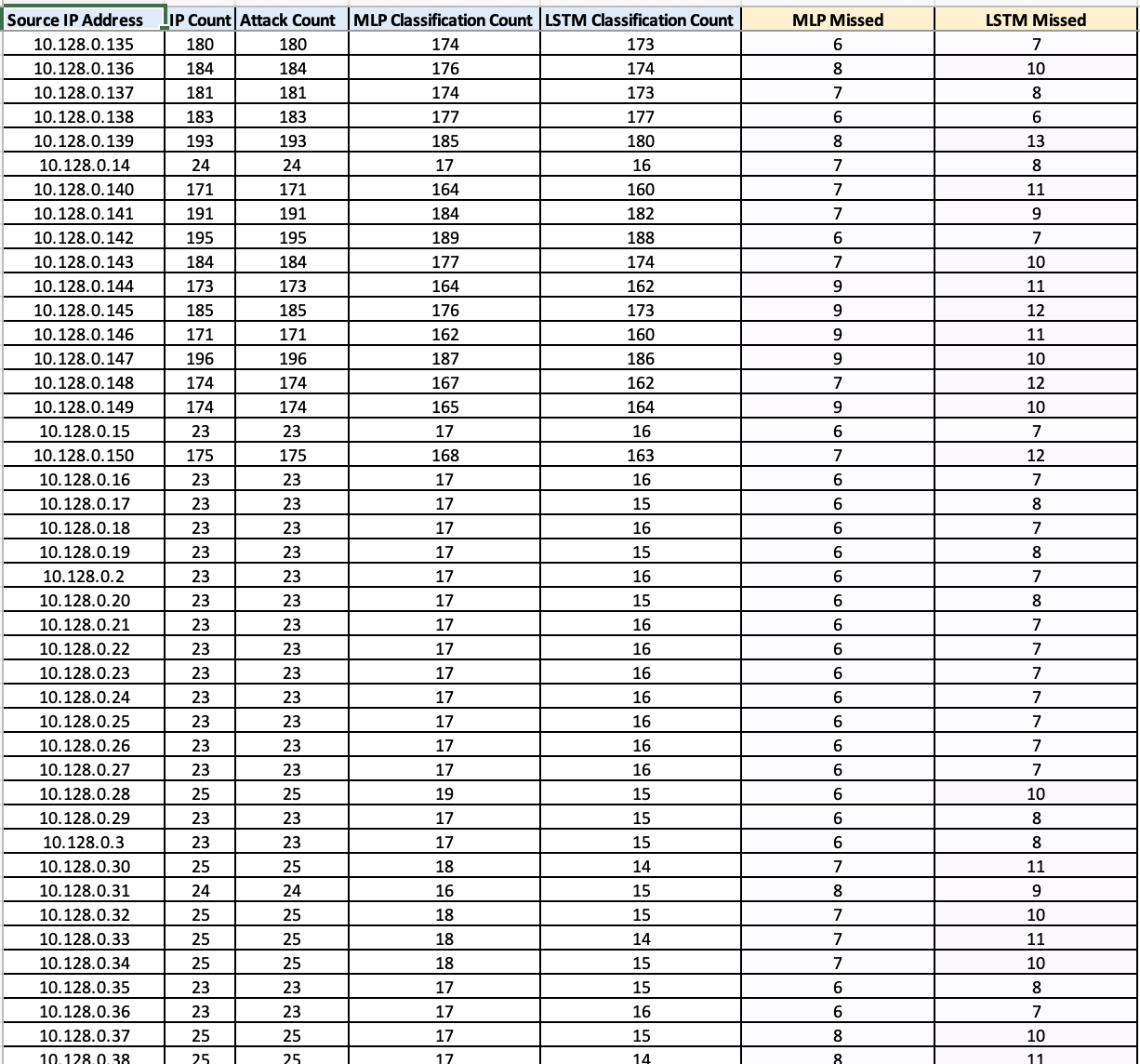}
  \caption{MLP and LSTM Model comparison. Red scale notes how many attack packets were missed by each mode. Yellow scale notes how many true benign packets were classified as attacks by the models.}
\end{figure}

\begin{figure}[H]
  \centering
  \includegraphics[scale=0.6]{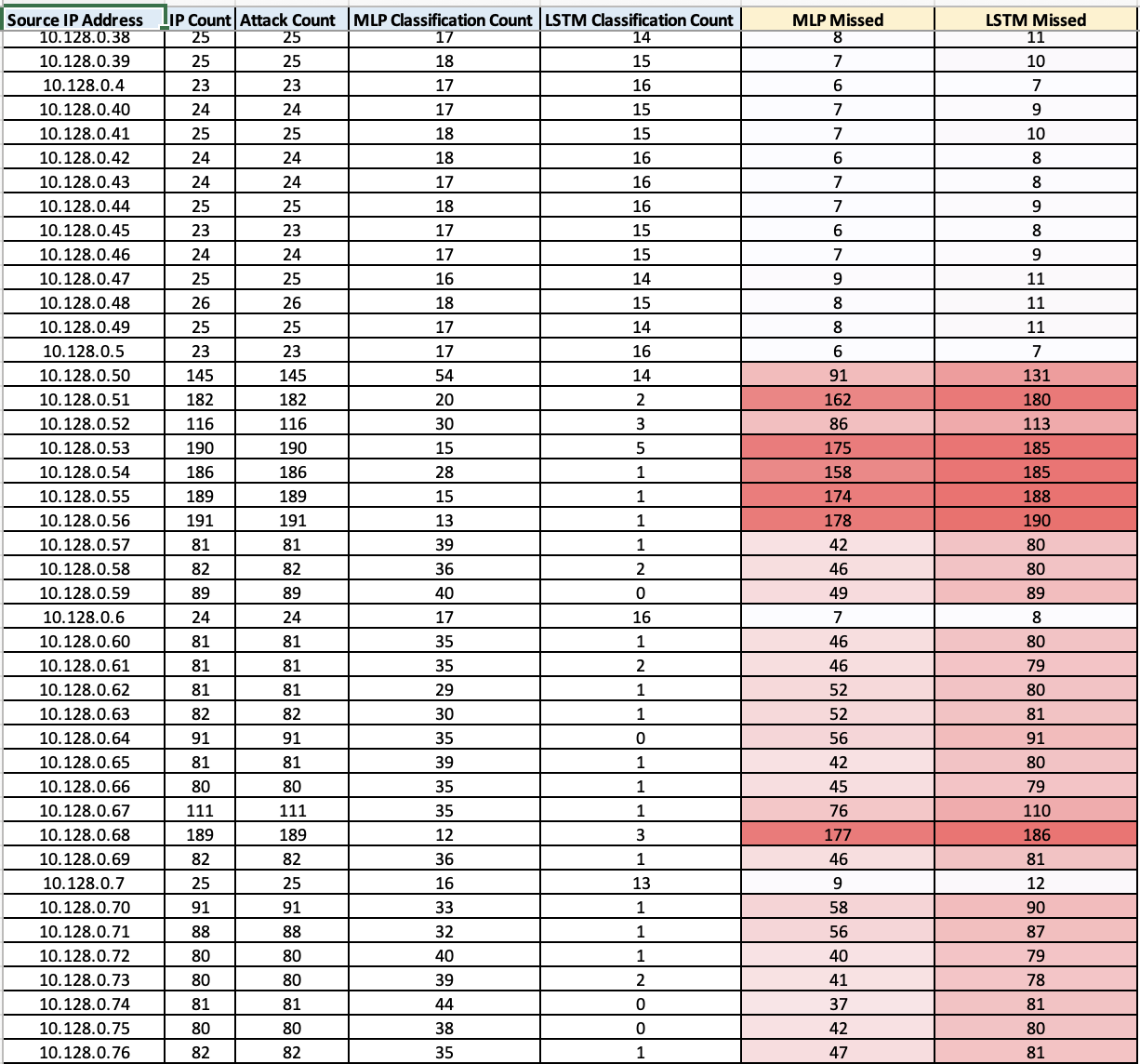}
  \caption{MLP and LSTM Model comparison. Red scale notes how many attack packets were missed by each mode. Yellow scale notes how many true benign packets were classified as attacks by the models.}
\end{figure}

\begin{figure}[H]
  \centering
  \includegraphics[scale=0.6]{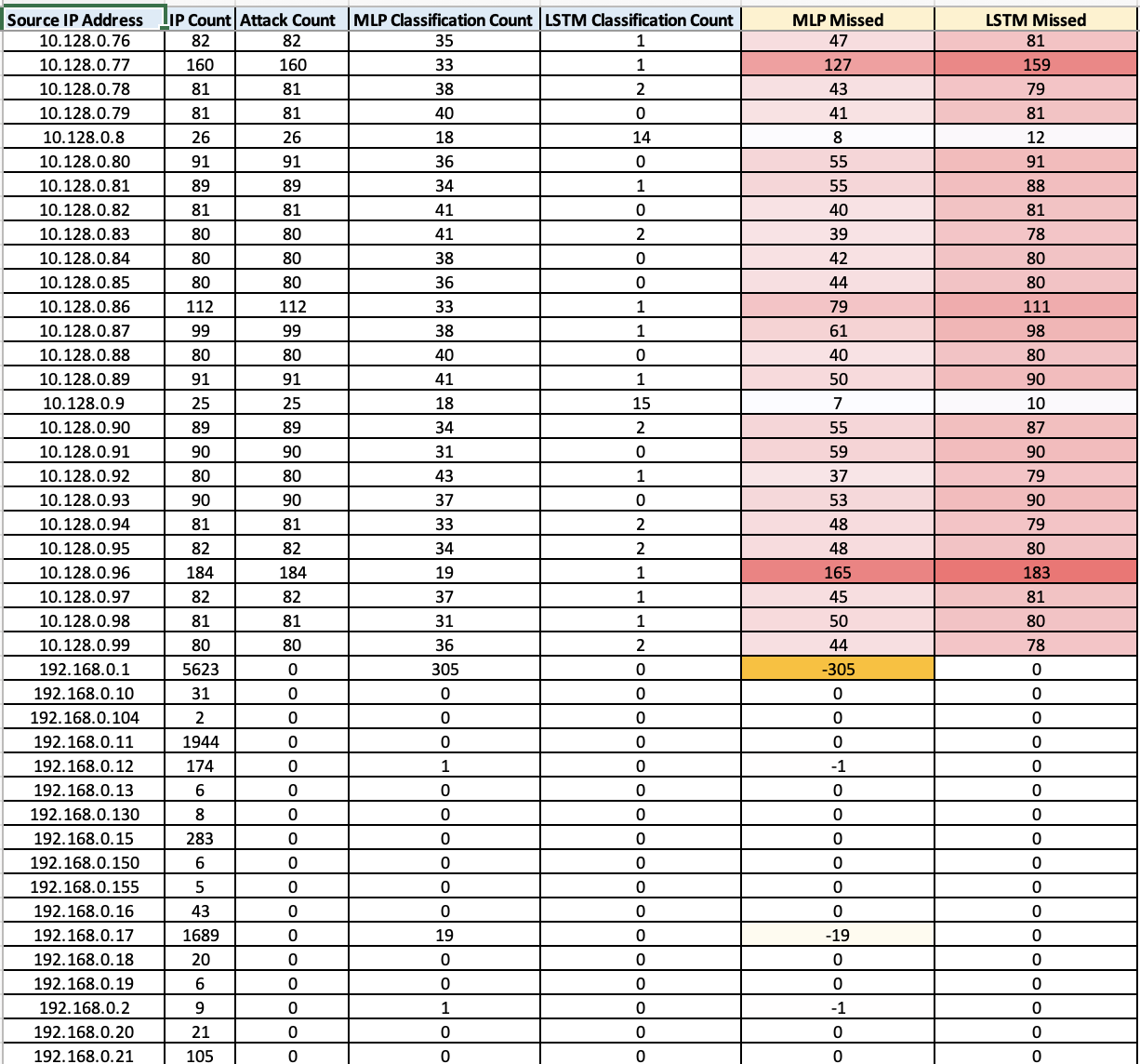}
  \caption{MLP and LSTM Model comparison. Red scale notes how many attack packets were missed by each mode. Yellow scale notes how many true benign packets were classified as attacks by the models.}
\end{figure}

\begin{figure}[H]
  \centering
  \includegraphics[scale=0.6]{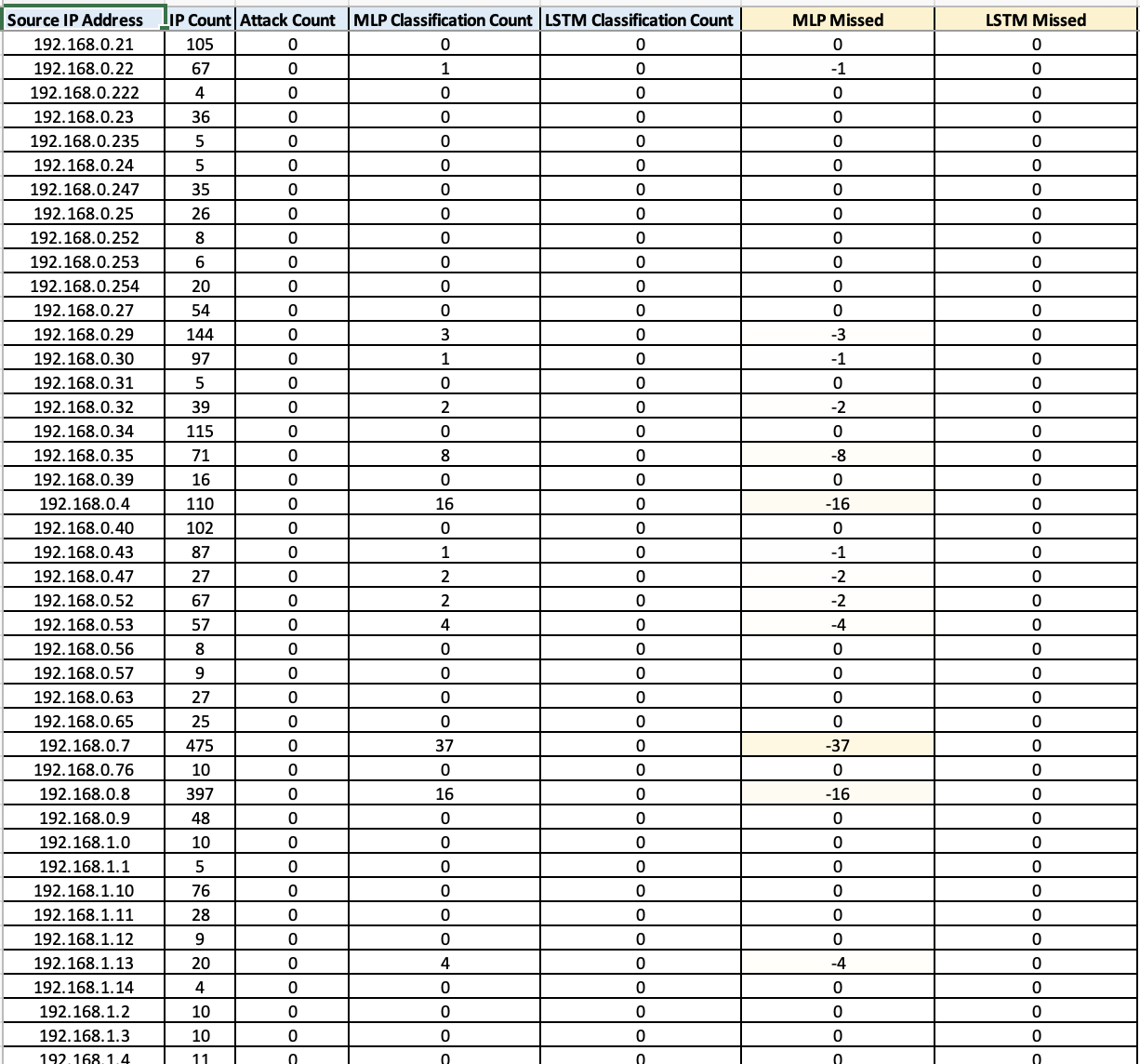}
  \caption{MLP and LSTM Model comparison. Red scale notes how many attack packets were missed by each mode. Yellow scale notes how many true benign packets were classified as attacks by the models.}
\end{figure}

\begin{figure}[H]
  \centering
  \includegraphics[scale=0.8]{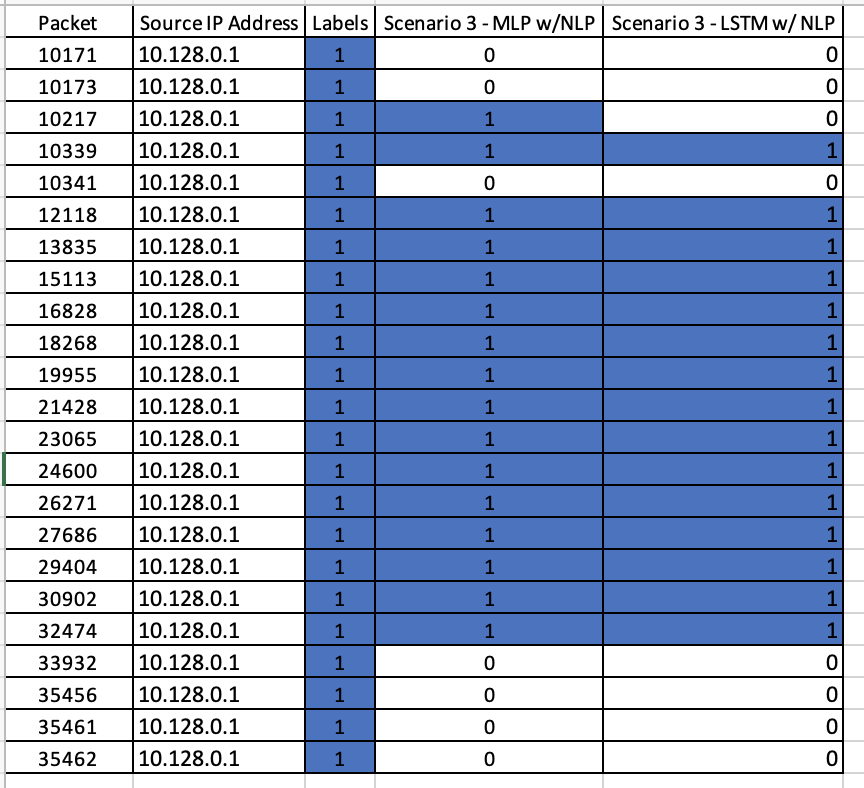}
  \caption{Timing results on a shorter attack}
\end{figure}

\begin{figure}[H]
  \centering
  \includegraphics[scale=0.8]{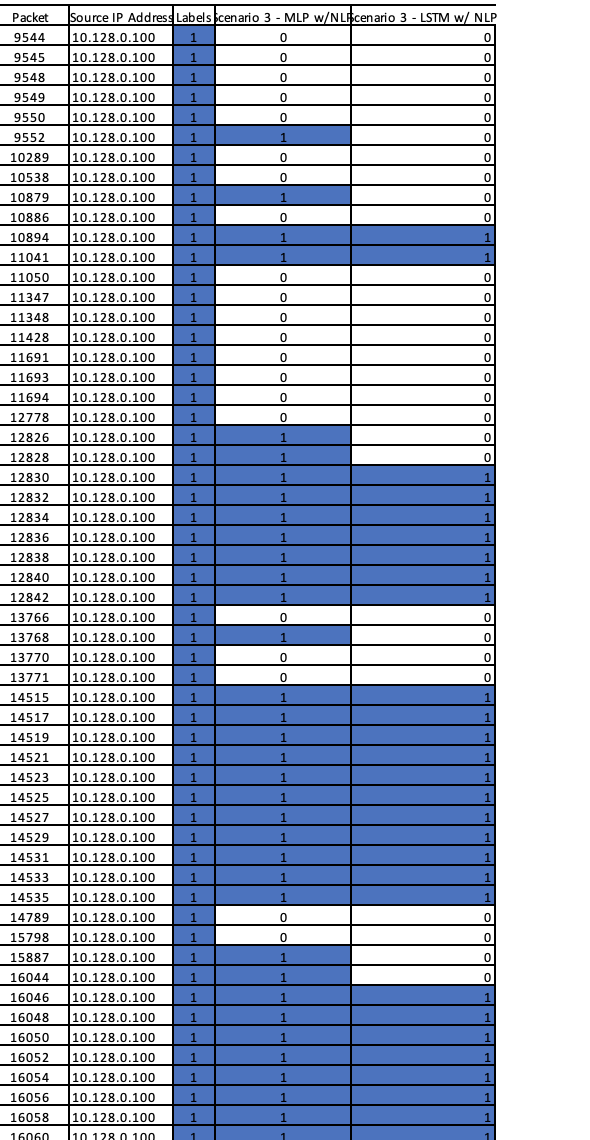}
  \caption{Timing results on a longer attack}
\end{figure}

\begin{figure}[H]
  \centering
  \includegraphics[scale=0.6]{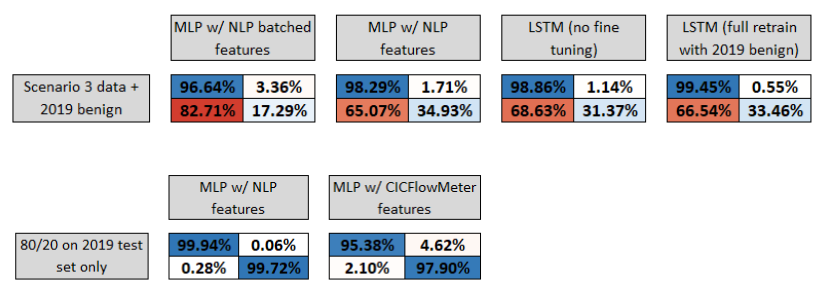}
  \caption{Final Evaluation Results on 2019 Dataset}
\end{figure}

\end{document}